\documentclass[superscriptaddress,aps, prl, reprint, nofootinbib]{revtex4-2}

\usepackage{orcidlink}
\usepackage{enumerate}
\usepackage{graphicx}

\usepackage{amsmath,amssymb}
\usepackage{dcolumn}
\usepackage{bm}
\usepackage[capitalise]{cleveref}
\usepackage{relsize}
\usepackage{aas_macros}
\usepackage{hyperref}

\usepackage{float} 

\newcommand{\ud}{\mathrm{d}}

\newcommand{\diff}{\mathrm{d}}

\colorlet{myurlcolor}{violet}

\hypersetup{
colorlinks=true, 
linkcolor=blue, 
citecolor=blue, 
filecolor=blue, 
urlcolor=blue, 
runcolor=blue
}

\definecolor{blue2}{cmyk}{1, 0.1, 0.1, 0.1}

\usepackage{colortbl}
\definecolor{lightgreen}{cmyk}{0.2, 0, 0.2, 0.2}
\definecolor{lightgray2}{cmyk}{0.1,0.1,0,0.1}
\definecolor{Red2}{RGB}{214, 39, 40}
\definecolor{Blue2}{RGB} {31, 119, 180}
\definecolor{Orange2}{RGB}{255, 127, 14}
\definecolor{Green2}{RGB}{44, 160, 44}
\definecolor{greyish2}{rgb}{.96,.96,.96}
\definecolor{lightblue}{rgb}{0.45, 0.59, 0.89}

\definecolor{pyBlue}{RGB}{31, 119, 180}
\definecolor{pyRed}{RGB}{214, 39, 40}
\definecolor{pyGreen}{RGB}{44, 160, 44}
\definecolor{pyBlue2}{RGB}{0, 111, 237}
\definecolor{pyRed2}{RGB}{224, 52, 36}
\definecolor{Mathematica1}{rgb}{0.368417, 0.506779, 0.709798}
\definecolor{Mathematica2}{rgb}{0.880722, 0.611041, 0.142051}

\def\beq{\begin{equation}}
\def\eeq{\end{equation}}

\begin{document}

\title{Dark Matter Mounds: towards a realistic description of dark matter overdensities around black holes}

\author{Gianfranco Bertone\,\orcidlink{0000-0002-6191-1487}}
\affiliation{Gravitation Astroparticle Physics Amsterdam (GRAPPA),
University of Amsterdam, Amsterdam, 1098 XH, Netherlands}

\author{A. Renske A. C. Wierda\,\orcidlink{0000-0002-9569-2745}}
\affiliation{Gravitation Astroparticle Physics Amsterdam (GRAPPA),
University of Amsterdam, Amsterdam, 1098 XH, Netherlands}
\affiliation{Department of Physics, KTH Royal Institute of Technology, The Oskar Klein Centre, AlbaNova, SE-106 91 Stockholm, Sweden}

\author{Daniele Gaggero\,\orcidlink{0000-0003-3534-1406}}
\affiliation{INFN Sezione di Pisa, Polo Fibonacci, Largo B. Pontecorvo 3, 56127 Pisa, Italy}

\author{Bradley J. Kavanagh\,\orcidlink{0000-0002-3634-4679}}
\affiliation{Instituto de F\'isica de Cantabria (IFCA, UC-CSIC), Avenida de Los Castros s/n, 39005 Santander, Spain}

\author{Marta Volonteri\,\orcidlink{0000-0002-3216-1322}}
\affiliation{Gravitation Astroparticle Physics Amsterdam (GRAPPA),
University of Amsterdam, Amsterdam, 1098 XH, Netherlands}
\affiliation{Institut d'Astrophysique de Paris, Sorbonne Universit\'e, CNRS, UMR 7095, 98 bis bd Arago, 75014 Paris, France}

\author{Naoki Yoshida\,\orcidlink{0000-0001-7925-238X}}
\affiliation{Department of Physics, The University of Tokyo, 7-3-1 Hongo, Bunkyo, Tokyo 113-0033, Japan}
\affiliation{Kavli Institute for the Physics and Mathematics of the Universe (WPI), UT Institute for Advanced Study, The University of Tokyo, Kashiwa, Chiba 277-8583, Japan}

\begin{abstract}
Dark matter overdensities around black holes can be searched for by looking at the characteristic imprint they leave on the gravitational waveform of binary black hole mergers. Current theoretical predictions of the density profile of dark matter overdensities are based on highly idealised formation scenarios, in which black holes are assumed to grow adiabatically from an infinitesimal seed to their final mass, compressing dark matter cusps at the center of galactic halos into very dense `spikes'. These scenarios were suitable for dark matter indirect detection studies, since annihilating dark matter cannot reach very high densities, but they fail to capture the dark matter distribution in the innermost regions where the gravitational wave signal is produced. We present here a more realistic formation scenario where black holes form from the collapse of  supermassive stars, and follow the evolution of the dark matter density as the supermassive star grows and collapses to a black hole. We show that in this case dark matter forms shallower `mounds', instead of `spikes', on scales comparable with the size of the supermassive stars originating them. We discuss the implications for the detectability of these systems.
\end{abstract}

\maketitle

{\bf Introduction.} The nature of Dark Matter (DM) remains one of the greatest unsolved mysteries of cosmology and particle physics~\cite{Bertone:2004pz,Bertone:2016nfn,Planck:2018vyg}. If DM is made of cold and collisionless particles then its density around Black Holes (BHs) will inevitably be higher than on average in the Universe, and possibly much higher. The presence of dense DM ``spikes'' around BHs~\cite{Gondolo:1999ef} would modify the dynamics of BH binaries, and induce a potentially detectable dephasing in the  gravitational waveforms of extreme- and intermediate-mass ratio inspirals (EMRIs and IMRIs, respectively)~\cite{Eda:2013gg,Eda:2014kra,Macedo:2013qea,Barausse:2014tra,Yue:2019ozq,Kavanagh:2020cfn,Becker:2021ivq,Speeney:2022ryg,Cardoso:2022whc,Nichols:2023ufs,Karydas:2024fcn,Kavanagh:2024lgq}. If current and future GW observatories detect such a dephasing, it will be possible to reconstruct information about the density of the spike~\cite{Cardoso:2019rou,Coogan:2021uqv,Cole:2022ucw}, to distinguish from other environmental effects~\cite{Cole:2022yzw}, and to use this information to distinguish between different DM models~\cite{Hannuksela:2019vip}. BH environments are therefore extremely promising probes of both the presence and nature of DM~\cite{Bertone:2019irm}. 
Whether such gravitational wave (GW) signatures are observable depends crucially on the abundance and properties of DM spikes in our Universe. DM spikes around supermassive BHs at the centers of galaxies~\cite{Gondolo:1999ef} might be significantly disrupted by mergers with other BHs of comparable mass, and might be depleted by interactions with dense stellar cusps~\cite{Ullio:2001fb,Merritt:2002vj,Bertone:2005hw}. Large DM overdensities are instead likely to persist around BHs that do not experience major BH-BH mergers, as well as around intermediate-mass BHs~\cite{Bertone:2005xz} and primordial BHs~\cite{Mack:2006gz,Adamek:2019gns,Boudaud:2021irr}. 

{\bf Black Hole formation.} 
Previous studies have assumed that DM spikes form when a BH grows adiabatically in mass from an infinitesimally small seed, 
following the prescription of Gondolo \& Silk~\cite{Gondolo:1999ef} (hereafter GS).
This approximation was suitable for the specific case of self-annihilating dark matter studied in that paper, since the physics of black hole formation and growth happens on scales smaller than the so-called annihilation plateau, a core of approximately constant density produced by DM annihilation~\cite{Bertone:2005xz}. 
But it fails to capture the time evolution of the density profiles of generic non-annihilating candidates in the innermost regions where deteactable GWs may be produced. 

The formation of massive astrophysical BHs is still poorly understood~\cite{Volonteri:2021sfo}, but in many astrophysical scenarios they form from the collapse of stellar objects, be they Pop III~\cite{Madau:2001sc}; very massive stars~\cite{PortegiesZwart:2004ggg}; or supermassive stars~\cite{Bromm:2002hb,Hosokawa:2013mba,2021ApJ...915..110W}. In the case of Pop III stellar collapse, the location of the stars may not coincide with the center of the DM halo except perhaps for PopIII.1~\cite{2019MNRAS.483.3592B}, while in the case of very massive stars generated by stellar collisions, the presence of other stars orbiting around the center can affect the ability of DM to attain high densities~\citep{Bertone:2005hw}. Instead, the BH likely forms at the center of the DM halo in direct collapse black hole scenarios, where supermassive star formation is mediated by extreme dissociating radiation and the cloud collapses spherically, e.g.~\citep{2014MNRAS.443.1979L,2018MNRAS.475.4104C}.

{\bf Dark matter {\it \bf Mounds}.} In this letter, we present the first self-consistent description of the formation of a DM overdensity around an astrophysical BH. Instead of the adiabatic growth of an infinitesimal mass seed, we follow here the evolution of the DM distribution in three phases: 
    i) adiabatic growth of a (super)massive stellar object at the centre of a DM halo;
 ii) instantaneous collapse of the (super)massive star to a BH; and 
 iii) adiabatic growth of the BH from the initial seed to its final mass. 
We demonstrate here that this process leads to a DM overdensity which is  time-dependent and shallower than DM spikes, which we generically refer to as DM `Mounds'. While this result has important consequences for DM-induced dephasing of GW signals, it also places such calculations on a more realistic footing and opens the way for population-level studies of DM spikes. 

\begin{figure}[t]
\hspace{-0.2cm}
\includegraphics[width=0.49\textwidth]{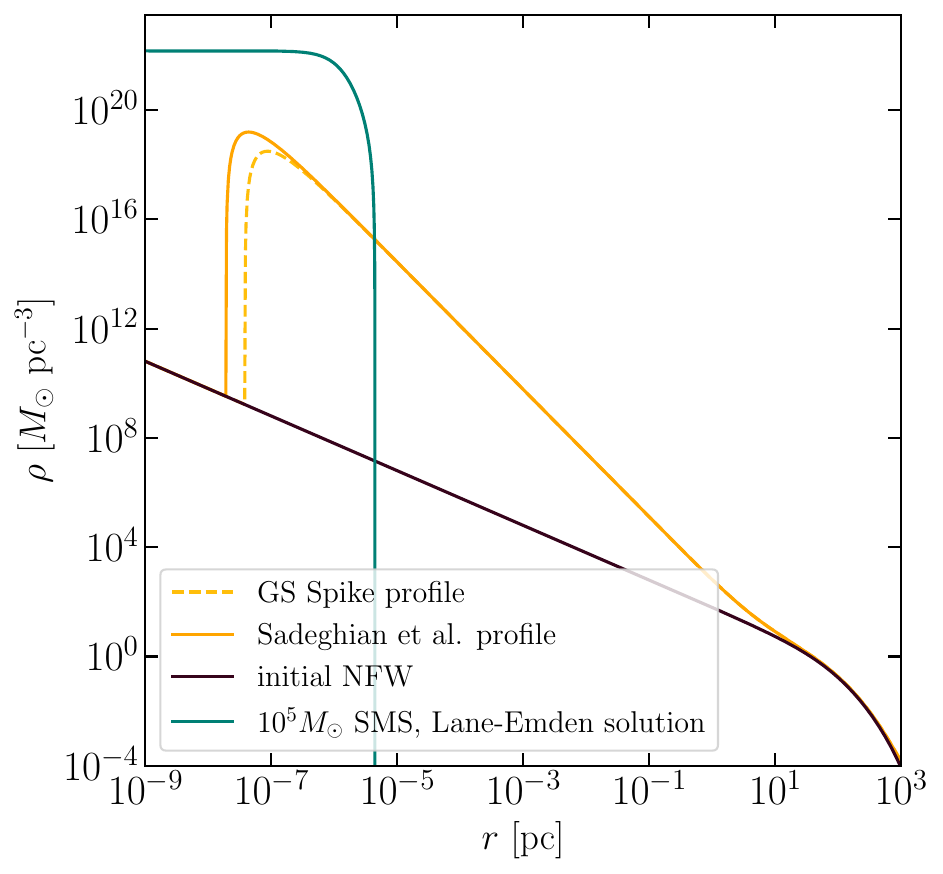}
\caption{Density profile of a $10^5\, M_{\odot}$ polytropic Supermassive Star (SMS) with polytropic index $n=3$, compared to the density profile of the dark matter halo hosting it, with virial mass $M_{\rm vir} = 10^7 \,M_{\odot}$ and concentration $c=3$ at redshift $z_{\rm f}=15$. The yellow lines corresponds to the non-relativistic spike profile of Ref.~\cite{Gondolo:1999ef} and the relativistic profile of Ref.~\cite{Sadeghian:2013laa}.}
\label{fig:SMS}
\end{figure}

{\bf Supermassive star formation.} 
We assume a set of fiducial values for the parameters describing the supermassive star (SMS) and DM density profile,
which are motivated by the results of recent cosmological simulations of radiation-driven direct-collapse BH formation~\citep{Chon16}.
Our formalism can be easily generalised to other scenarios, as long as the SMS forms at the center of the DM halo. We consider in particular a spherically symmetric SMS with mass $M_{\rm SMS}=10^5 M_{\odot}$, with central density $\rho_c=1$ g cm$^{-3}$~\citep{Hosokawa:2013mba}.
The interior of an SMS is supported by radiation pressure,
and while detailed calculations of stellar evolution show that an accreting SMS 
has a core-envelope structure with a bloated, high-entropy atmosphere \citep{Hosokawa:2013mba,Umeda16}, the core structure is well approximated by an $n=3$ polytrope.

The formation timescale for a $10^5 \, M_{\odot}$ SMS growing at a rate $\dot{M}=0.1 \,M_{\odot} \,\rm{yr}^{-1}$ is $t_{\rm{growth}}=M/\dot{M}=10^6$ years. This is longer than the typical dynamical timescale at the radius of the SMS, $t_{\rm{dyn}}=\sqrt{(3 \pi)/(32 G \bar{\rho})}\sim 10^3$ years (see \cref{fig:SMS}). The timescale for hydrodynamical processes is even shorter, meaning that the SMS is expected to maintain its core-envelope structure during accretion.
The growth timescale of $10^6$ years is then longer than any other relevant dynamical timescales.

{\bf Initial DM profile.}  We assume that the SMS forms at the center of a DM halo with virial mass $M_{\rm vir} = 10^7\, M_{\odot}$ at redshift $z_{\rm f}=15$. The DM density profile before SMS formation is described by a Navarro-Frenk-White (NFW)  profile~\cite{Navarro:1995iw}. The concentration $c(M_{\rm vir},z)$ of the NFW profile is derived from a semi-analytic model based on an extended Press-Schechter theory for the halo mass accretion history, and calibrated with numerical simulations~\cite{Correa:2015dva}. For our fiducial values of the virial mass and redshift described above, $c = 3$. 
Under these approximations, the matter distribution in the initial system is completely described.  
We show in Fig.~\ref{fig:SMS} the density profile of the SMS and of the dark matter profile hosting it. We also show for comparison the GS spike profile which would result from the adiabatic growth of a $10^5\,M_\odot$ black hole at the centre of the halo.

{\bf Adiabatic growth.} In order to compute the dark matter density after SMS formation, we need to generalise the GS spike formation formalism.\footnote{The local velocity dispersion in the potential of the SMS does not typically exceed a few percent of the speed of light, so we here safely use the non-relatistic formalism of GS.} Assuming spherical symmetry, the angular momentum $L$ and the radial action $I(\mathcal{E},L)$ are adiabatic invariants. We denote with $\cal{E}$ the relative energy per unit mass ${\cal E} = \Psi(r) - v^2/2$, where
$\Psi(r) = \phi_0 - \phi(r)$ is the positive-definite relative gravitational potential, so that all particles with ${\cal E} > 0$ are on bound orbits. We can write the radial action as
\begin{equation}
    \label{eqn:radial_action}
    I({\cal E}, L) = \frac{1}{\pi} \int_{r_\mathrm{min}}^{r_\mathrm{max}}{ \ud r \; v_r(r, \mathcal{E}, L)}\,,
\end{equation}
where the radial speed is given by 
\begin{equation}
v_r = \sqrt{2\Psi(r) - 2 {\cal E}-\frac{L^2}{r^2}} \,,
\end{equation}
and where $r_\mathrm{min}$ and $r_\mathrm{max}$ are the pericentre and apocentre of the orbit, given by the solutions to $v_r = 0$.

The phase space distribution of DM particles is described by the distribution function $f(\mathcal{E},L)$, which is conserved under adiabatic changes of the gravitational potential: $f_f({\cal E}_f,L) = f_i({\cal E}_i,L)$. The initial distribution function is evaluated following the Eddington inversion procedure, assuming a spherically symmetric and isotropic NFW halo~\cite{BinneyAndTremaine,Lacroix:2018qqh}. The initial and final energies of particles are related through conservation of the radial action.
The initial and final radial actions are evaluated on a grid of $(\mathcal{E}_i, L)$ and $(\mathcal{E}_f, L)$; in the initial configuration we fix $\Psi(r)$ equal to the potential of the NFW profile, while in the final configuration we also include the contribution of the SMS, computed numerically.  Given the conservation of $I$ and $L$ under adiabatic conditions, we determine $\mathcal{E}_i$ by finding, through interpolation, the value which satisfies $I_i({\cal E}_i,L)=I_f({\cal E}_f,L)$ for each $\mathcal{E}_f$ and $L$. 
With this, we can evaluate $f_f({\cal E}_f,L)$ and thus reconstruct the final density after adiabatic growth of the star as:
\begin{equation}
     \rho_f(r) = \frac{4 \pi}{r^2} \int_0^{{\cal E}_f^{\rm max}} \mathrm{d}{\cal E}_f 
     \int_{0}^{L_f^{\rm max}}\mathrm{d}L_f \frac{L_f}{v_r} f_f({\cal E}_f,L_f)\,.
\label{eq:finprof}
\end{equation}
The maximum angular momentum is given by $L_f^\mathrm{max} = \sqrt{2 r^2 \left (\Psi(r) - \mathcal{E}_f\right)}$ and the maximum energy is given implicitly by $\mathcal{E}_f^\mathrm{max} = \Psi(r)$.

{\bf Collapse.} 
The accreting supermassive star collapses to a BH after about $10^5$ year. The exact timing of the gravitational collapse depends on the gas mass accretion rate~\citep{Umeda16,Haemmerle18}, but it occurs roughly 
near the end of the hydrogen burning phase.
After the onset of implosion, either after hydrogen exhaustion or when triggered by general relativistic instability, the star contracts and collapses very rapidly. We assume that the entire mass of the SMS collapses to form the BH. 
The sudden change in gravitational potential modifies the orbits of individual dark matter particles. The angular momentum of the initial and final orbits will be the same, but the final energy $\mathcal{E}_f$ is related to the initial energy $\mathcal{E}_i$ as $\mathcal{E}_f = \mathcal{E}_i + \Delta \Psi(r)$, where $\Delta\Psi$ is the difference between the final potential due to the BH ($m_\mathrm{BH} = 10^5\,M_\odot$) and the initial potential due to the supermassive star.
\begin{figure}[tb] 
\centering
\includegraphics[width=0.49\textwidth]{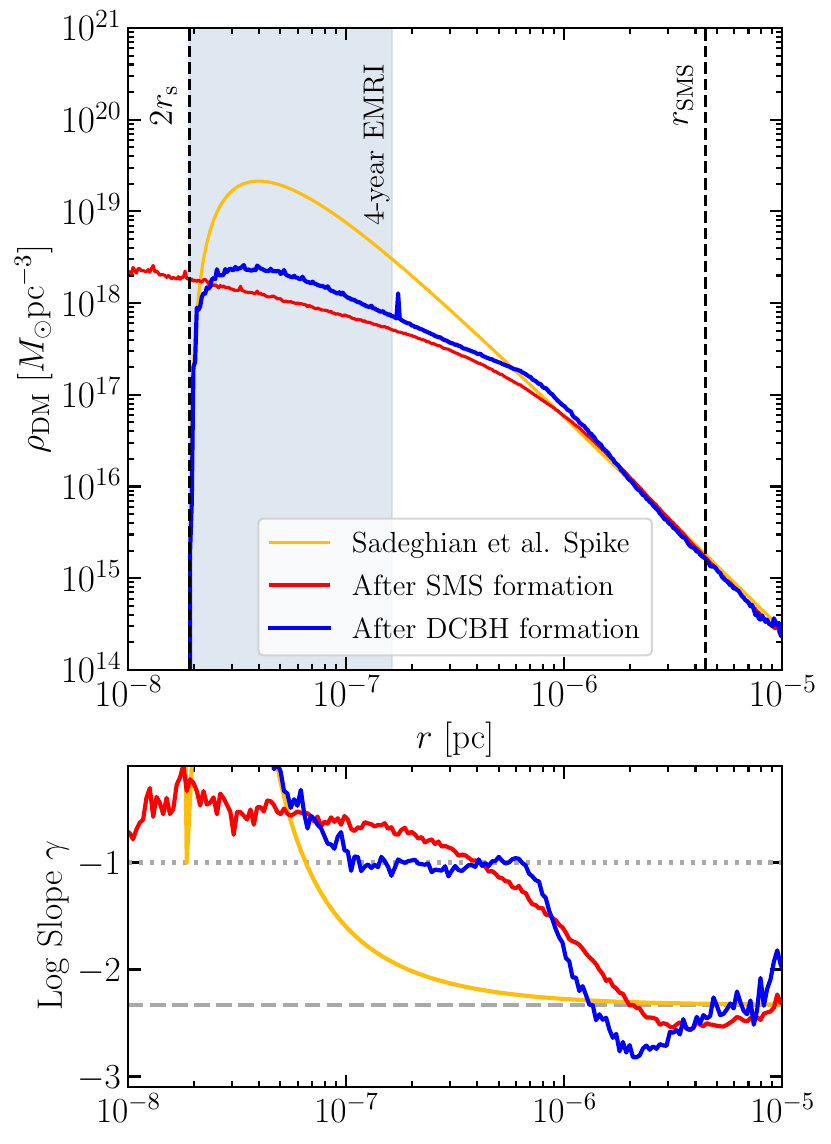}
\caption{\textit{Top panel:} Density profile of dark matter after the formation of the supermassive star (red), and after the collapse to a black hole (blue). For comparison, we show the relativistic spike solution of Ref.~\cite{Sadeghian:2013laa} (yellow). The vertical lines denote the cutoff radii $2r_s$, the separation of the EMRI binary 4 years before merger, and the radius of the SMS. \textit{Bottom panel:} Logarithmic slope of the profiles. The horizontal lines correspond to the slopes $-7/3$ and $-1$, see text for further details.}
\label{fig:blunt}
\end{figure}

Following Ref.~\cite{Ullio:2001fb}, for an orbit specified by energy $\mathcal{E}$ and angular momentum $L$, the probability of finding after collapse the particle at a radius $r$ is given by:
\begin{equation}
P(r|\mathcal{E}, L)\,\diff r=\frac{2}{T(\mathcal{E}, L)}\frac{1}{\tilde{v}_{r}(r,\mathcal{E}, L)}\,\diff r  \,,
\label{eq:prob}
\end{equation}
where $T(\mathcal{E}, L) = 2 \int_{r_\mathrm{min}}^{r_\mathrm{max}} \tilde{v}_r^{-1}\,\diff r$ is the period of the orbit. Post-collapse, we use the radial velocity in the Schwarzschild metric~\cite{Sadeghian:2013laa}:
\begin{equation}
\label{eq:v_r_schwarzschild}
       \tilde{v}_r^2 = c^2 - 2 \mathcal{E}- \left(c^2 - \frac{2 G M_\mathrm{BH}}{r}\right)\left(1 + \frac{L^2}{r^2 c^2}\right)\,.
\end{equation}
Ref.~\cite{Ullio:2001fb} computed the final density profile in the case of initially circular orbits, whereas we are interested here in the more complex case of an initial DM distribution arising from the adiabatic contraction of an initial NFW profile following the SMS growth. To obtain the final density profile, we thus need to integrate the probability in \cref{eq:prob} over the appropriate initial distribution function:
\begin{equation}
     \rho_f(r) = \frac{1}{4 \pi r^2} \int \diff^3 \mathbf{r}_i \,\diff^3 \mathbf{v}_i  \frac{2 f_i(\mathcal{E}_i,L)}{T(\mathcal{E}_f,L) \,\tilde{v}_{r}(r,\mathcal{E}_f, L)}\,,
\end{equation}
where we recall that $\mathcal{E}_f = \mathcal{E}_i + \Delta \Psi(r_i)$. 
As appropriate, this equation reduces to Eq.~(11) in Ref.~\cite{Ullio:2001fb}, in the special case of circular orbits. Changing variables from $\mathbf{v}_i$ to ${\cal E}_i$ and $L$, we find: 
\begin{equation}
     \rho_f(r) = \frac{8 \pi}{r^2} \int \,\diff r_i \,\frac{L\,\diff L\,\diff {\cal E}_i}{v_{r}(r_i,\mathcal{E}_i, L)\,}  \frac{f_i(\mathcal{E}_i,L)} { T(\mathcal{E}_f,L) \,\tilde{v}_{r}(r,\mathcal{E}_f, L)}\,.
     \label{eq:rhobh}
\end{equation}

In practice, we evaluate the integral in \cref{eq:rhobh} with a Monte Carlo procedure. We draw $N = 10^6$ samples of initial orbits $\left\{{\cal E}_{i,k},L_{k}\right\}$, with associated weights $w_k$, proportional to $L_k \,f_i(\mathcal{E}_{i,k},L_{k})$. For each sample, we also draw an initial radius $r_{i,k}$ from the probability distribution in \cref{eq:prob}, in order to calculate ${\cal E}_{f,k} = {\cal E}_{i,k} + \Delta \Psi(r_{i, k})$. The final density is then evaluated by summing over the samples:
\begin{equation}
    \rho_f(r) = \frac{8\pi}{r^2}\times \frac{1}{N} \sum_k \frac{w_k}{T(\mathcal{E}_{f, k},L_k) \,\tilde{v}_{r}(r,\mathcal{E}_{f,k}, L_k)}\,.
\end{equation}
We apply cuts on $(\mathcal{E}_f, L)$ following the prescription in Ref.~\cite{Sadeghian:2013laa}, in order to remove orbits which are unstable or are immediately captured by the BH. Full details of these cuts (and a comparison with the fully relativistic calculation) are given in the Supplementary Material.

{\bf Results.} 
In \cref{fig:blunt}, we show the dark matter density profile after the formation of the supermassive star (solid red), and after the collapse to a black hole (solid blue). For comparison, we also show the spike profile expected from the adiabatic growth of a $10^5\,M_\odot$ BH (``Sadeghian et al.\ Spike"), without the intermediate stage of SMS formation. 
The grey shaded region in \cref{fig:blunt} highlights the range of radii probed by an extreme mass-ratio inspiral (EMRI, $m_1 = 10^5 \,M_\odot$, $m_2 = 10 \,M_\odot$), starting 4 years before the merger.

We see that the density after SMS formation is significantly lower than the spike profile of Refs.~\cite{Gondolo:1999ef,Sadeghian:2013laa}, in the region where the gravitational wave signal is expected to peak. 
Note also that the density profile extends below the Schwarzschild radius of the black hole, since collapse has not taken place yet. When the SMS collapses to a BH, the profile steepens again, due to the increased potential experienced by particles inside the star, but remains below the spike solution, except in a small range of radii ($r \sim 10^{-6}\,\mathrm{pc}$), where the profile slightly exceeds the standard spike solution. This `excess' can be understood in terms of mass conservation. The central density is lower than in the standard spike solution (because adiabatic contraction is less efficient at $r \ll r_\mathrm{SMS}$), meaning that the density at radii just below $r \sim r_\mathrm{SMS}$ must be larger.  

In the lower panel of  \cref{fig:blunt}, we show the logarithmic slope of the DM density profiles at each stage. Overall, following the evolution of the dark matter density profile through the formation of the SMS, and then through its collapse, leads to a profile which matches the spike solution ($\gamma = -7/3$) outside the radius of the SMS but is significantly shallower inside.
 We find this behavior for a wide range of values for the total DM mass and the SMS mass. For more massive SMSs, we find that the shallower inner region is larger, as expected.
The behaviour of the mounds can be understood qualitatively, as we demonstrate for circular orbits in the Supplementary Material. Conservation of DM mass and angular momentum in the initial phase of adiabatic growth of the SMS leads to a slope shallower than $-1$, while the subsequent collapse to a BH should lead to a slope of $-5/4$, further flattened to a slope of around $\gamma = -1$ by the effects of capture by the BH.

\begin{figure}[tb] 
\centering
\hspace{-0.3cm}
\includegraphics[width=0.49\textwidth]{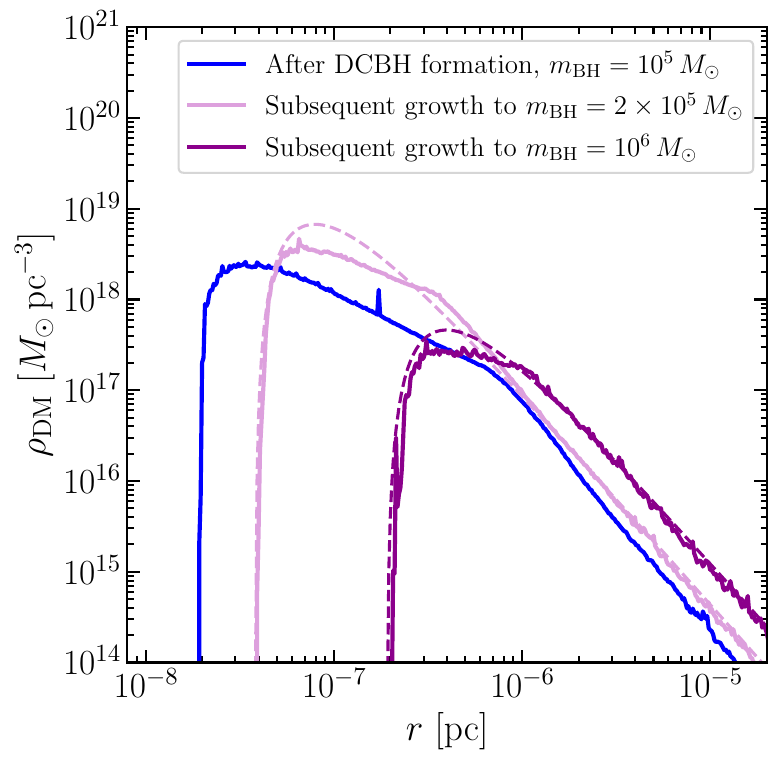}
\vspace{-0.5cm}
\caption{Density profile of dark matter after the formation of the direct collapse black hole ($m_\mathrm{BH} = 10^5\,M_\odot$, blue line) and after the subsequent adiabatic growth of the BH by a factor of 2 (light purple) and a factor of 10 (dark purple). For comparison, we show the corresponding relativistic spike profile following the formalism of  Ref.~\cite{Sadeghian:2013laa} (dashed lines), assuming the adiabatic growth of a BH with the given mass from a seed of negligible mass at the centre of an NFW halo, without the intermediate stages described above.}
\label{fig:growth}
\end{figure}

In \cref{fig:growth}, we show the DM density profile following the subsequent adiabatic growth of the DCBH from $10^5\,M_\odot$ to $2 \times 10^5\,M_\odot$ (light purple) and to $10^6 \,M_\odot$ (dark purple). We also show a comparison with the Sadeghian et al.\ relativistic spike profile for these two final BH masses~\cite{Sadeghian:2013laa}.  As we increase the final BH mass, we find that the spike normalisation at large radii grows as $m_\mathrm{BH}^{2/3}$, as expected from the GS formalism, up to a factor of $\sim 5$ (for growth up to $10^6\,M_\odot$).
Increasing the final mass also increases the Schwarzschild radius of the BH, leading to a cut-off in the spike at larger and larger radii. With growth by a factor of 2, the flattened inner density profile is still visible in the spike. For growth by a factor of 10, however, the outer power-law slope appears to persist down to the cut-off. In this case, the final DM density matches that expected for the adiabatic growth of an infinitesimal seed down to a few Schwarzschild radii, with a mild suppression of the density below this radius.

{\bf Conclusions.} A detailed understanding of the dark matter profiles in EMRI/IMRI systems is essential to measure and interpret the dephasing of the gravitational waveform induced by dark matter overdensities. We have shown that previous calculations of the dark matter profile, while suitable for indirect dark matter searches, fail to properly characterise the density in the region of interest for gravitational wave observations. We have presented a 3-step formation scenario that follows the evolution of the dark matter profile from an initial NFW profile (\cref{fig:SMS}), through the formation of the progenitor SMS and its collapse to a BH (\cref{fig:blunt}, top panel), to the final growth of the BH to its final mass (\cref{fig:growth}). 

In order to compute the evolution of dark matter after collapse, we have introduced a new method, based on a Monte Carlo procedure, which generalises the results of Ref. \cite{Ullio:2001fb} to non-circular orbits, and includes the relativistic corrections to the inner cut-off from Ref. \cite{Sadeghian:2013laa}.
We have demonstrated that the DM profile around the BH is significantly shallower than previously thought on scales smaller than the SMS radius, right after collapse (\cref{fig:blunt}, bottom panel), leading to the formation of rounded peaks that we have dubbed `mounds' to discriminate them from the sharper `spikes' introduced in Ref.~\cite{Gondolo:1999ef}. If the BH grows substantially in mass from this initial seed, the combined effect of instantaneous collapse of the SMS to a BH, and the inclusion of relativistic effects, leads to a final profile that is similar to a `spike', but which may be flatter in the innermost regions.

Upcoming GW experiments like LISA \cite{Baker:2019nia}, DECIGO \cite{Kawamura:2020pcg} and TianQin \cite{TianQin:2015yph} are expected to precisely probe BH environments \cite{Cole:2022yzw,Baumann:2021fkf,Baumann:2022pkl,Derdzinski:2020wlw}. A fraction of the EMRI/IMRI events detectable with these experiments at low redshift would in particular carry the distinctive signature of the dark matter environment around the central BH \cite{Coogan:2021uqv}. This opens up the possibility to discriminate `mounds' from `spikes' by studying the dephasing of the gravitational waveforms, and thus to infer the formation and growth history of massive BHs.  Our 3-step formation scenario can be easily extended to other SMS masses and radial profiles, and other DM halo properties, as long as the SMS forms adiabatically at the center of DM halo. This approach can therefore be implemented in numerical and semi-analytical methods for the calculation of EMRI and IMRI rates, and their detectability with future interferometers. 

\textit{Acknowledgements.} GB gratefully acknowledges the Department of Physics of Columbia U. and the Italian Academy for Advanced Studies in America, where part of this work was carried out. D.G.~acknowledges support from the project ``Theoretical Astroparticle Physics (TAsP)'' funded by INFN.
BJK acknowledges funding from the \textit{Consolidaci\'on Investigadora} Project \textsc{DarkSpikesGW}, reference CNS2023-144071, financed by MCIN/AEI/10.13039/501100011033 and by the European Union ``NextGenerationEU"/PRTR.

\bibliography{main2}

\appendix
\section{Supplementary Material}
\subsection{Relativistic Density Profile}
Here, we present further details of how we incorporate relativistic effects in the analysis. While the energy before and after adiabatic growth are calculated non-relativistically, we include relevant relativistic effects in the final density profile through (a) the use of the radial velocity in the Schwarzschild metric in Eq.~\eqref{eq:v_r_schwarzschild}; and (b) cuts on the energy $\mathcal{E}$ and angular momentum $L$ to account for DM capture by the black hole. 

These cuts are described in Ref.~\cite{Sadeghian:2013laa} and exclude orbits which pass through the Schwarzschild radius, as well as unstable orbits which will eventually be captured. To first order in $\mathcal{E}$, the relativistic energy of the particles per unit mass can be written as $\mathcal{E}_\mathrm{rel} \approx c^2 - \mathcal{E}$. With this, the cuts amount to removing particles at a radius $r$ with:
\begin{align}
    \begin{split}
        L < L_c &\equiv 2 r_s c\left(1 - \frac{4\mathcal{E}}{c^2}\right)^{1/2}\\
        \mathcal{E} > \mathcal{E}_c &\equiv \frac{c^2}{2}\frac{r_s (r -  2r_s)}{(r^2 + 2 r r_s - 4 r_s^2)}\,.
    \end{split}
\end{align}
This correctly gives rise to a cut-off in the spike density profile at $r = 2 r_s$ as discussed in Ref.~\cite{Sadeghian:2013laa}.

\begin{figure}[tb] 
\includegraphics[width=0.49\textwidth]{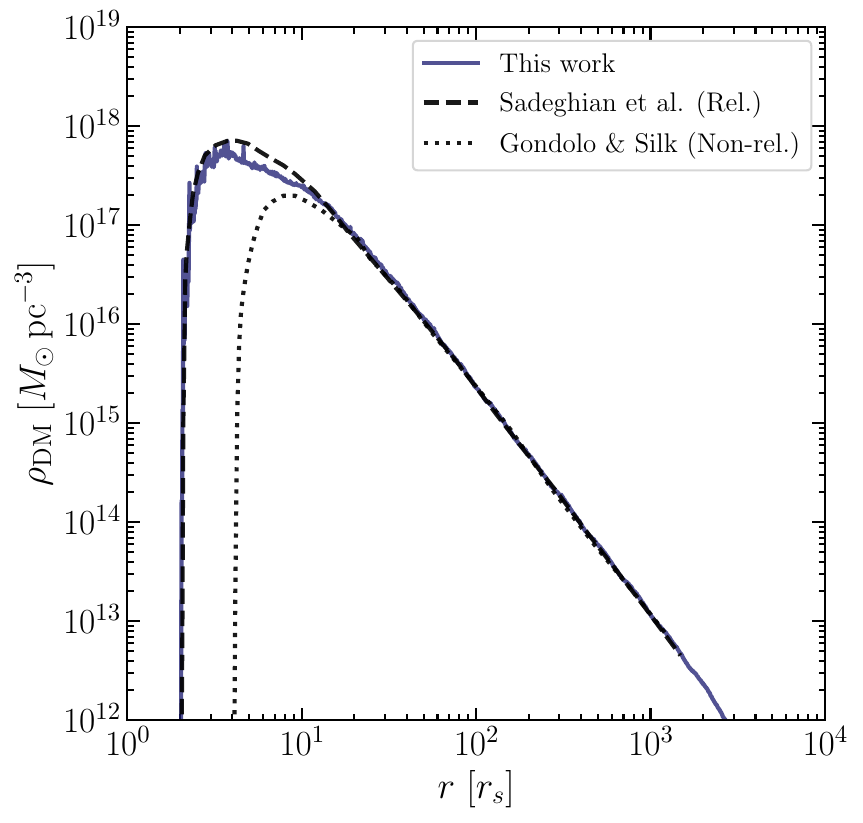}
\raggedright
\caption{Comparison of our formalism for adiabatic growth (from an infinitesimal seed) with the full relativistic formalism of Sadeghian et al.~\cite{Sadeghian:2013laa} and the non-relativistic formalism of Gondolo \& Silk~\cite{Gondolo:1999ef}. Here, we assume the adiabatic growth of a BH of mass $1.1\times10^6\,M_\odot$ at the centre of a Hernquist DM profile, as used in Ref.~\cite{Sadeghian:2013laa}. The dotted and dashed lines are taken from Fig.~3 of Ref.~\cite{Sadeghian:2013laa}.}
\label{fig:Sadeghian_Comparison}
\end{figure}

In order to verify that this prescription provides an accurate estimate of the density profile, we show in Fig.~\ref{fig:Sadeghian_Comparison} the non-relativistic spike of Gondolo \& Silk~\cite{Gondolo:1999ef} (dotted line); the fully relativistic spike of Sadeghian et al.~\cite{Sadeghian:2013laa} (dashed line); and the formalism including relativistic corrections we present here (solid line). In all cases, the spike is grown adiabatically from an infinitesimal seed (without the intermediate stages described in the main text). We see that our formalism matches very closely that of Sadeghian et al.\ down to $r \lesssim 10 r_s$. Below this radius, our result deviates by at most 30\% from the fully-relativistic result, though it provide a substantial improvement over the non-relativistic GS spike. We are therefore confident that the most important relativistic effects are included in the analysis, especially considering at below $\sim 5\, r_s$ the dynamics of IMRIs/EMRIs is expected to be dominated by GW emission, for which the precise DM spike density will be less relevant.

\subsection{Analytic Arguments}

This section provides detailed analytic arguments to explore the slopes of the density profiles of the DM mounds found in \cref{fig:blunt}. We first treat the adiabatic growth of the supermassive star, following by the subsequent instantaneous collapse of the star to form a BH. 

{\bf Adiabatic Growth.} Following Ref.~\cite{Quinlan:1994ed}, we examine the effects of the adiabatic growth, assuming circular orbits for the DM particles. Considering the orbital radius before and after the growth of the supermassive star (SMS), $r_i$ and $r_f$, conservation of DM mass gives:
\begin{align}
    \begin{split}
     \rho_i(r_i) \,r_i^2\,\diff r_i &\sim \rho_f(r_f) \,r_f^2\,\diff r_f  \\
     \Rightarrow r_i^{3 + \gamma_i} &\sim r_f^{3 + \gamma_f}\,,
    \end{split}
\end{align}
where $\gamma_i$ and $\gamma_f$ are the slopes of the initial and final power-law density profiles. 

Conservation of angular momentum gives:
\begin{align}
    \begin{split}
     r_i \,m_{\mathrm{enc},i}(r_i) &\sim r_f \,m_{\mathrm{enc},f}(r_f)  \\
     \Rightarrow r_i^{4 + \gamma_i} &\sim r_f^4\,\,.
    \end{split}
\end{align}
Here, $m_\mathrm{enc}(r)$ is the enclosed mass as a function of radius. In the final case, we have approximated the SMS as having a constant density core out to a radius $r_\mathrm{SMS}$, leading to an enclosed mass $m_\mathrm{enc}(r) \approx m_\mathrm{SMS}(r^3/r_\mathrm{SMS}^3)$ for $r < r_\mathrm{SMS}$. We have also assumed that the mass of the SMS dominates the DM mass at small radii.  

Solving for $\gamma_f$, we find:
\begin{equation}
\gamma_f = \frac{4 (3+\gamma_i)}{(4 +\gamma_i)} - 3\,,
\end{equation}
or $\gamma_f \approx -1/3$ for an initial NFW profile ($\gamma_i = -1$). In the innermost regions of the mound profile, where the DM orbits are completely contained within the SMS, we find that the slope does indeed tend to this value (see the numerical results in \cref{fig:blunt}). However, at larger radii $r \lesssim r_\mathrm{SMS}$, we find in our numerical results a steeper slope of $\gamma \sim -0.5$. The central density is enhanced by orbits whose apocentre lies outside (or close to) the SMS radius. As these orbits enter the SMS their orbital velocity is smaller than it would be in a point-potential with the same total mass. They therefore spend more time at this smaller radii, enhancing the central density and steeping the slope of the density profile. 

\begin{figure}[tb] 
\centering
\includegraphics[width=0.49\textwidth]{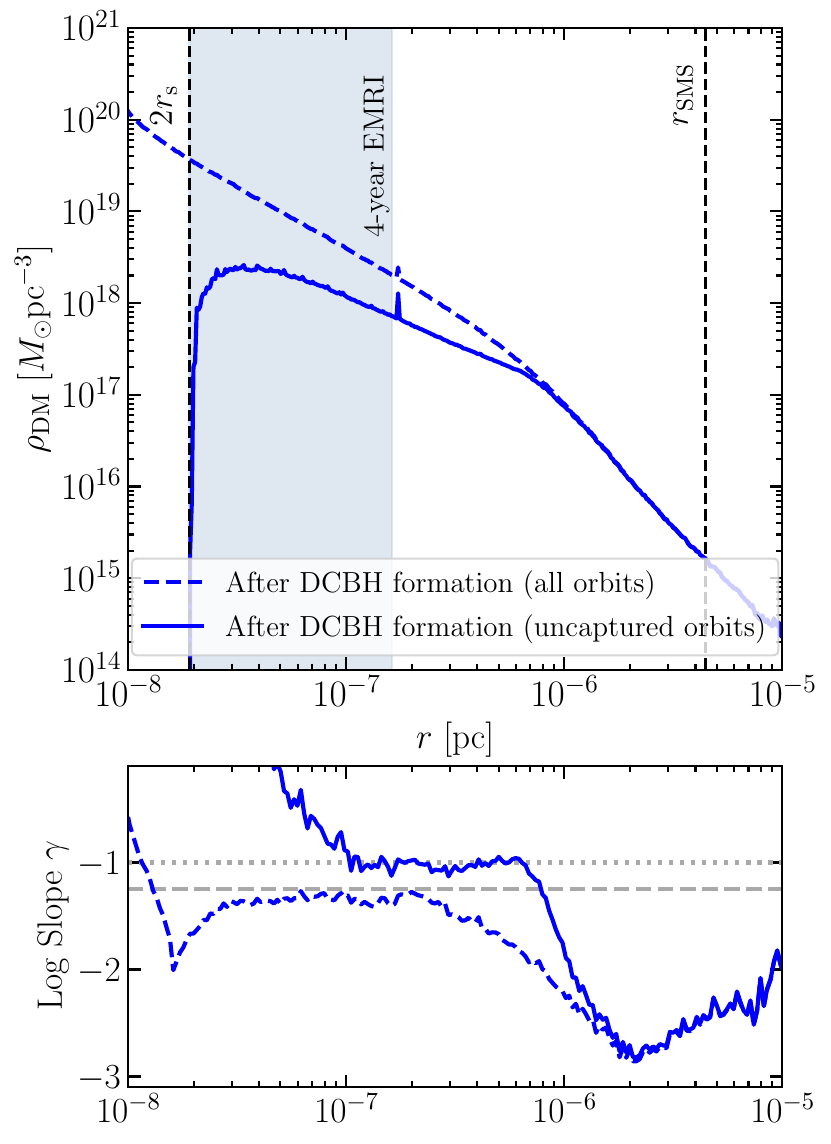}
\caption{Same as \cref{fig:blunt}, but comparing the density profile after DCBH formation with (solid blue) and without (dashed blue) accounting for the capture of orbits by the newly formed BH. Without accounting for capture, the inner logarithmic slope of the DM density profile is $-5/4$, in agreement with our analytic arguments. This is then flattened towards $-1$ when including the effects of capture.}
\label{fig:uncaptured}
\end{figure}

{\bf Instantaneous Collapse.}
For illustration, we will assume circular orbits for the DM particles before the collapse of the SMS to a BH. Following Ref.~\cite{Ullio:2001fb}, we can write the final density after collapse by integrating over the contributions of particles located at radius $r_0$ at the moment of collapse:
\begin{equation}
    \rho_f(r) = \frac{1}{r^2}\int_{r_\mathrm{min}}^{r_\mathrm{max}}\,\diff r_0 \,r_0^2 \rho_i(r_0)\frac{2}{T(r_0)}\frac{1}{v_r(r, r_0)}\,,
    \label{eq:instant_collapse}
\end{equation}
where $T(r_0)$ and $v_r(r, r_0)$ are the post-collapse orbital period and radial velocity of a DM particle which was initially at a radius $r_0$. This radial velocity can be written:
\begin{align}
\begin{split}
    v_r(r, r_0)= & \sqrt{\frac{2 G M_{\mathrm{BH}}}{r r_0}} \\
& \times \sqrt{\left(r_0-r\right)\left(1-\frac{m_\mathrm{SMS}\left(r_0\right)}{2 m_{\mathrm{BH}}}\left(1+\frac{r_0}{r}\right)\right)} \,,
\end{split}
\end{align}
where $m_\mathrm{enc}(r_0)$ is the enclosed mass of the SMS at a radius $r_0$. We again approximate the SMS as having a constant density core, and note that $m_\mathrm{SMS} = m_\mathrm{BH}$ after collapse. The limits of the integral in \cref{eq:instant_collapse} define the range of initial radii of orbits which pass through $r$ after collapse. These limits can be found from the zeros of $v_r(r, r_0)$ as a function of $r_0$, for which we obtain $r_\mathrm{min} = r$ and $r_\mathrm{max} \approx (2 r r_\mathrm{SMS})^{1/4}$.

\begin{figure*}[t]
    \centering
        \includegraphics[width=0.3\textwidth]{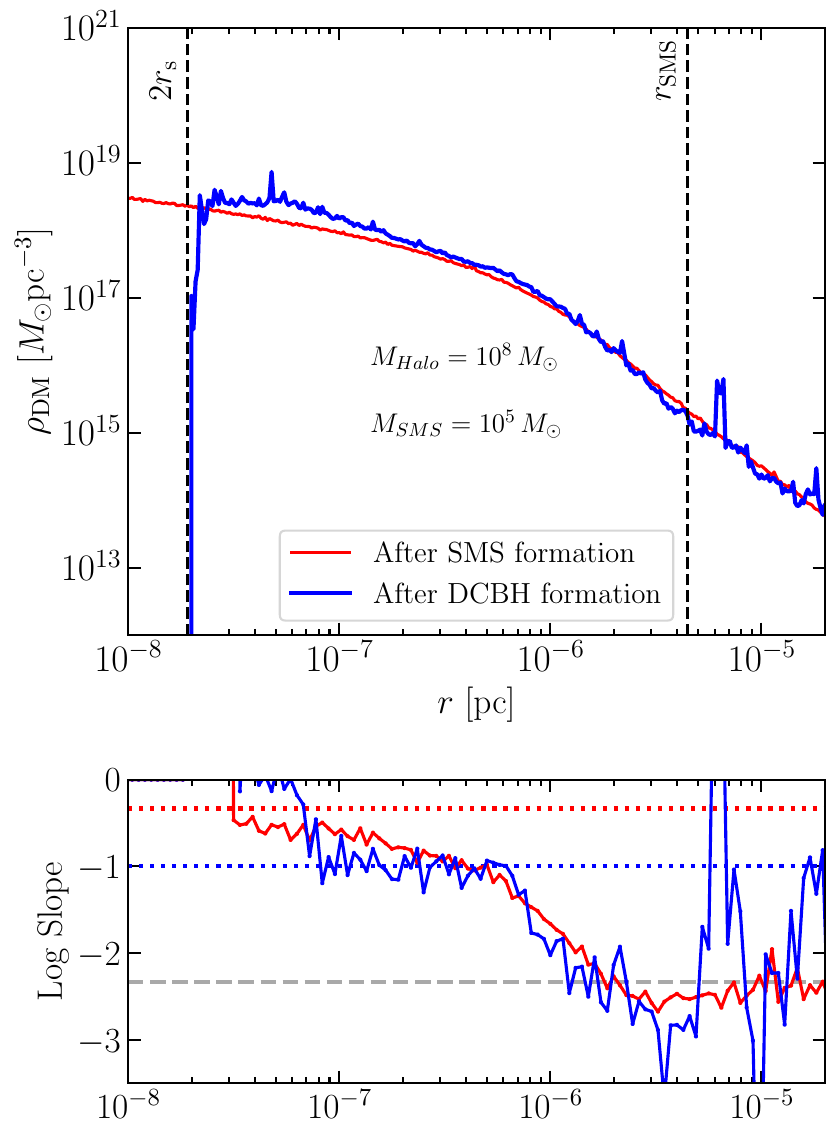}
        \includegraphics[width=0.3\textwidth]{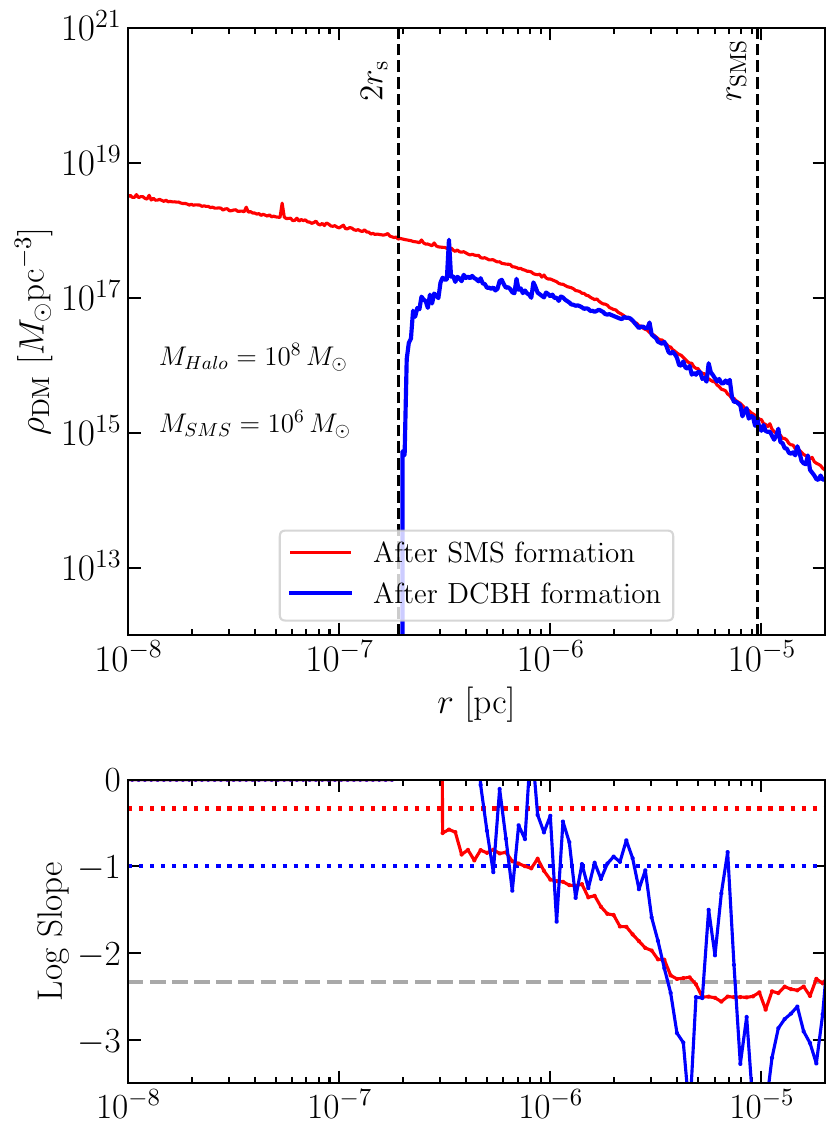}
        \includegraphics[width=0.3\textwidth]{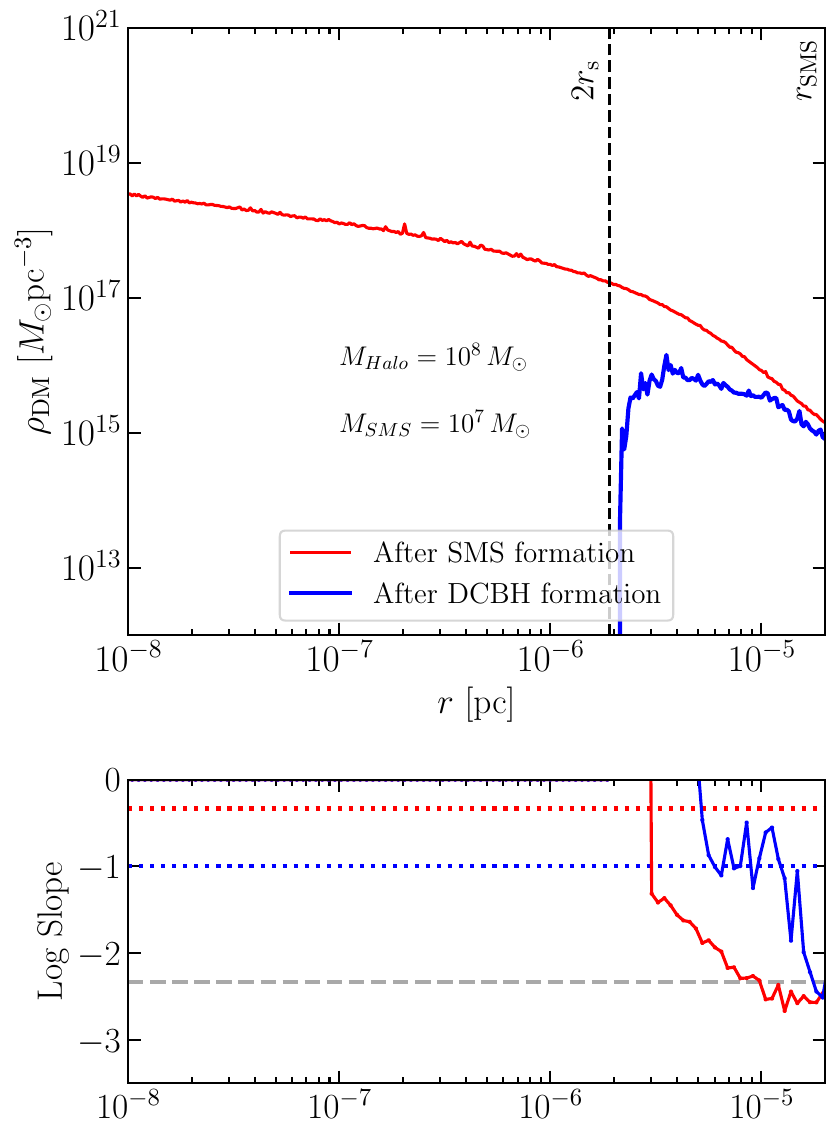}
    \caption{The top 3 panels show the density profile after SMS formation, and after DCBH formation, for a halo mass of $10^8$ $M_{\odot}$, and $M_\mathrm{SMS}$ equal to $10^5$ (left), $10^6$ (center) and to $10^7$ $M_{\odot}$ (right). The bottom panel shows the corresponding logarithmic slope as a function of radius.}
    \label{fig:SMS_massesC}
\end{figure*}
Noting that in the potential of the BH, the orbital period scales as $T(r_0) \sim r_0^{3/2}$ and writing the pre-collapse density profile as $\rho_i(r_0) \propto r_0^{\gamma_i}$, the final density profile scales as:
\begin{align}
    \rho_f(r) \sim \frac{r_\mathrm{max}^{3/2 + \gamma_i}}{r^{3/2}}\int_{x}^1  \frac{x_0^{1/2 + \gamma_i} \,\diff x_0}{\sqrt{(1 - \frac{x}{x_0})\left(1 - x_0^3 x\left(1 + \frac{x_0}{x}\right)\right)}}\,,
    \label{eq:rho_f_scaling}
\end{align}
where we have defined the dimensionless quantities $x = r/r_\mathrm{max}$ and $x_0 = r_0/r_\mathrm{max}$. For small $r$, the lower limit tends to $0$ and the dimensionless integral becomes roughly independent of $r$. Recalling that $r_\mathrm{max} \sim r^{1/4}$ and that $\gamma_i \approx -0.5$, we thus expect $\rho_f(r) \sim r^{-5/4}$ after collapse. 

In \cref{fig:uncaptured}, we show the DM density profile after DCBH formation which we obtain in our full numerical analysis, when the effects of capture by the BH are ignored (dashed blue line). As we see in the lower panel, the inner density profile in this case matches very well the slope of $-5/4$ derived above. This power-law is further flattened once we include the effects of capture (solid blue line). We assume that a DM particle will be captured if the pericentre $r_\mathrm{peri}$ of its orbit is less than $r_\mathrm{cap} = 2 r_s$. This requirement equates roughly to requiring $r_0 > (2 r_\mathrm{cap}r_\mathrm{SMS}^3)^{1/4}$, thus altering the lower limit of the integral in \cref{eq:rho_f_scaling}, from $x$ to $(2 r_\mathrm{cap}r_\mathrm{SMS}^3)^{1/4}/r_\mathrm{max} \sim (r_\mathrm{cap}/r)^{1/4}$. The lower limit of the dimensionless integral now grows as $r \rightarrow 0$ and the integral now scales roughly as $(1 - (r_\mathrm{cap}/r)^{1/4})$ for sufficiently small $r$. This leads to a smooth flattening from a slope of $-5/4$ towards zero (eventually turning over and giving a cut-off at  $r = r_\mathrm{cap}$). The behaviour shown in \cref{fig:blunt} is more complicated; in that case, the initial velocities are not circular and so the behaviour of $r_\mathrm{peri}$ with initial radius is more complicated.

\subsection{Robustness tests}

In this section, we demonstrate the robustness of our results against variations in halo mass and SMS mass values. We show the results for two different scenarios:

\begin{itemize}
\item We keep the SMS mass fixed to a value of $10^5$ $M_{\odot}$ and vary the halo mass from $10^6$ to $10^8$ $M_{\odot}$. 
\item We keep the halo fixed to a value of $10^8$ $M_{\odot}$ and vary $M_\mathrm{SMS}$ from $10^5$ to $10^7$ $M_{\odot}$.
\end{itemize}

In all cases we keep the concentration constant ($c = 3$), given the negligible variation of this parameter with the halo mass at the redshift of interest \cite{Correa:2015dva}. 

We find very little dependence of our results on the halo mass, as expected given that the region that is relevant for our treatment is dominated by the potential of the central object. 
Regarding variations of the SMS mass, we observe an enlargement of the flatter inner region for larger values of this parameter, as expected. 
We show in Fig.~\ref{fig:SMS_massesC}  a set of plots that illustrate this result, showing both the change in the profile and the change in the behavior of the log slope as a function of radius, for 3 different values of the SMS mass.

\end{document}